\documentclass[aip,jcp,reprint]{revtex4-1}

\usepackage{graphicx}
\usepackage{graphics}
\usepackage{amsfonts}
\usepackage{amsmath}
\usepackage{color}
\usepackage{enumerate}
\usepackage{multirow}
\usepackage{gensymb}
\usepackage{hyperref}

\newcommand{\revision}[1]{{#1}}

\begin{document}

\title{Ion-ion correlations across and between electrified graphene layers}
\author{Trinidad Mendez-Morales}
\affiliation{Maison de la Simulation, CEA, CNRS, Univ. Paris-Sud, UVSQ, Universit\'e Paris-Saclay, 91191 Gif-sur-Yvette, France} 
\affiliation{Sorbonne Universit\'e, CNRS, Physico-chimie des \'electrolytes et nanosyst\`emes interfaciaux, PHENIX, F-75005 Paris, France}
\affiliation{R\'eseau sur le Stockage \'Electrochimique de l'\'Energie (RS2E), FR CNRS 3459, France}

\author{Mario Burbano}
\affiliation{Maison de la Simulation, CEA, CNRS, Univ. Paris-Sud, UVSQ, Universit\'e Paris-Saclay, 91191 Gif-sur-Yvette, France} 
\affiliation{Sorbonne Universit\'e, CNRS, Physico-chimie des \'electrolytes et nanosyst\`emes interfaciaux, PHENIX, F-75005 Paris, France}
\affiliation{R\'eseau sur le Stockage \'Electrochimique de l'\'Energie (RS2E), FR CNRS 3459, France}

\author{Matthieu Haefele}
\affiliation{Maison de la Simulation, CEA, CNRS, Univ. Paris-Sud, UVSQ, Universit\'e Paris-Saclay, 91191 Gif-sur-Yvette, France} 

\author{Benjamin Rotenberg}
\affiliation{Sorbonne Universit\'e, CNRS, Physico-chimie des \'electrolytes et nanosyst\`emes interfaciaux, PHENIX, F-75005 Paris, France}
\affiliation{R\'eseau sur le Stockage \'Electrochimique de l'\'Energie (RS2E), FR CNRS 3459, France}

\author{Mathieu Salanne}
\email{mathieu.salanne@sorbonne-universite.fr}
\affiliation{Maison de la Simulation, CEA, CNRS, Univ. Paris-Sud, UVSQ, Universit\'e Paris-Saclay, 91191 Gif-sur-Yvette, France} 
\affiliation{Sorbonne Universit\'e, CNRS, Physico-chimie des \'electrolytes et nanosyst\`emes interfaciaux, PHENIX, F-75005 Paris, France}
\affiliation{R\'eseau sur le Stockage \'Electrochimique de l'\'Energie (RS2E), FR CNRS 3459, France}

\begin{abstract}

When an ionic liquid adsorbs onto a porous electrode, its ionic arrangement is deeply modified due to a screening of the Coulombic interactions by the metallic surface and by the confinement imposed upon it by the electrode's morphology. In particular, ions of the same charge can approach at close contact, leading to the formation of a superionic state. The impact of an electrified surface placed between two liquid phases is much less understood. Here we simulate a full supercapacitor made of the 1-butyl-3-methylimidazolium hexafluorophosphate and nanoporous graphene electrodes, with varying distances between the graphene sheets. The electrodes are held at constant potential by allowing the carbon charges to fluctuate. Under strong confinement conditions, we show that ions of the same charge tend to adsorb in front of each other across the graphene plane. These correlations are allowed by the formation of a highly localized image charge on the carbon atoms between the ions. They are suppressed in larger pores, when the liquid adopts a bilayer structure between the graphene sheets. These effects are qualitatively similar to the recent templating effects which have been reported during the growth of nanocrystals on a graphene substrate.

\end{abstract}

\maketitle

\section{Introduction}

Economic growth is directly coupled with the increase of energy consumption. In this context, society is facing a double energy challenge in the 21st century: a global transition from fossil fuels to renewable energies motivated not only by the limit on the supply but also by the environmental impacts, and the development of new devices with an enhanced ability to store energy. With this in mind, it is not surprising that supercapacitors and batteries have attracted considerable attention from research agencies and industry due to their different properties and benefits for different applications. They differ in the way they store energy: Whereas supercapacitors use a reversible adsorption of the ions on the electrode surfaces with no charge transfer, batteries rely on a reversible redox reaction that occurs in the bulk of the electrode. As a consequence, supercapacitors have a much higher charge and discharge rate and can survive many more cycles than batteries. Thus, supercapacitors provide a much greater power density, but they are still not able to store the same amount of energy as the best-performing batteries, such as Li-ion~\cite{Clarke-2015, Barsukov-2003, Conway-1999}.  

In the last years, tremendous research efforts have been devoted to increasing the energy density of supercapacitors while maintaining their high power capability and their long cycling life \cite{Chen-2010, Mullen-2013, Huang-2015}. Taking into account that the stored energy in a supercapacitor can be calculated as $E=\frac{CV^{2}}{2}$ ($C$ being the total capacitance of the device and $V$ the cell voltage), two approaches can be followed to reach the objective of improving their overall performance: either to develop new electrode materials~\cite{frackowiak2013a,simon2008a,li2016c,zhu2016a} with increased capacitance or to design novel electrolytes with wider potential windows~\cite{brandt2012a,ray2016a,schutter2016a} to enhance the operating voltage (and theoretical studies also suggest a synergistic enhancement of the two quantities~\cite{lee2016a,kondrat2016a}). 

Concerning the electrolytes, they strongly affect the performance of supercapacitors and many eyes are fixed on increasing their stability under applied voltage. For this purpose, room temperature ionic liquids (ILs) can be considered as very promising candidates, since these 'designer solvents' are very well-known for exhibiting some unique properties such as high thermal stability, high electrochemical stability, low vapor pressure and non-toxicity \cite{MacFarlane-2014, Hussey-2007}. 
 
When it comes to increasing the capacitance, it has been previously shown, both experimentally and computationally, that the interaction between the electrode material and the electrolyte plays a fundamental role in optimizing the adsorption of the ions~\cite{fedorov2014a,salanne2016a,forse2016a}. For example, a big step forward was taken with the perception of the pore size effect and the ion desolvation under confinement made by Chmiola \textit{et al.}\cite{Simon-2006} when analyzing tetraethylammonium tetrafluoroborate (TEA-BF$_4$) in acetonitrile (ACN) using carbide-derived carbons (CDCs). An anomalous increase of the capacitance when the ion size matches the pore size was not only observed in traditional electrolytes \cite{Simon-2008}, but also employing ILs \cite{Simon-2008B, Ania-2006}. These findings broke with the traditional view that mesoporous electrodes ($>2\;\mathrm{nm}$) were necessary to allow the ions to enter the pores with their solvation shells intact \cite{Endo-2001}, and they drew the attention of the computational research community to gain deeper insight into the electrolyte environment at electrified surfaces~\cite{kirchner2013a,breitsprecher2015a,ivanistsev2014a,mendezmorales2014a} and under nanoconfinement \cite{Qiao-2011B, Kim-2010, Bedrov-2013, Vatamanu-2015B}. Molecular dynamics (MD) simulations employing realistic structural models of CDC electrodes demonstrated the better efficiency of nanoporous electrodes over the planar ones\cite{Merlet-2012B}, because they avoid the occurrence of overscreening effects and they allow a denser packing of the ions inside the pores due to the electronic screening by the pore walls of the ions' Coulombic interactions. This latter effect is known as ``superionic state'' and it was first reported by Kondrat and Kornyshev\cite{Kondrat-2011A,Kondrat-2011}. Vatamanu \textit{et al.}\cite{Bedrov-2013B} also showed a noticeable improvement in the capacitive storage when designing the electrodes with carbon single-chain segments, due to their high content of atomically rough and curved surfaces. Moving to ILs confined in slit-like pores, several groups \cite{Feng-2011, Jiang-2011} used simulations to predict oscillations of the capacitance as a function of the pore size. Further work insisted on the importance of pore length~\cite{breitsprecher2017a} and on the necessity to account for the pore size distribution to interpret the experimental data \cite{Kornyshev-2012}.

Based on these findings, many experimental groups have proposed graphene-based electrode materials to enhance supercapacitor performances~\cite{stoller2008a,liu2010a,yoo2011a,lian2017b}. In principle, it should provide the largest accessible surface among carbon materials since the ions can approach from both sides of the graphene plane. However the main difficulties consisted in i) avoiding the restacking of the layers when densifying the material and ii) providing a pathway for the ions to access the whole material. These difficulties were overcome by developing new synthesis methods. In particular, a supercapacitor made of the 1-ethyl-3-methylimidazolium bis(trifluoromethane)sulfonimide (EMI-TFSI) electrolyte
and holey graphene electrodes yielded a capacitance value of 45~F/g~\cite{han2014b}, 
while the same IL inside annealed microwave
exfoliated graphene oxide (MEGO) reached 130 F/g~\cite{tsai2013a}. The performances of these devices
were further expanded through chemical activation: for example the activated MEGO can store up to 
200~F/g~\cite{zhu2011a}. The wide range of experimentally measured values is due to
the variability in the structure of the materials, but it calls for further simulation studies to rationalize the results. Many works have simulated slit-pores, which neglects the fact that adjacent layers of liquid which are located on the two sides of a graphene sheet may interact with each other. With the aim of clarifying this aspect, here we carry out MD simulations of perforated nanoporous graphene in the IL 1-butyl-3-methylimidazolium hexafluorophosphate (BMIM-PF$_6$). In particular, we analyze the local structure of the ions inside the pores and we show that in the case where the IL is highly confined, significant correlations between two adjacent layers of liquid arise. This feature is reminiscent of
the ``graphene transparency'' previously reported 
for the wettability of water~\cite{rafiee2012a} and
the growth of nanocrystals~\cite{chae2017a} on supported graphene,
and it may have important consequences for charge storage in graphene-based supercapacitor 
electrodes.

\section{Simulation details}
MD simulations of pure BMIM-PF$_6$ confined between graphene-based electrodes were carried out within an \textit{NVE} ensemble at an average temperature of $400\;\mathrm{K}$, which was chosen due to the high viscosity of this IL at room temperature ($109.2\;\mathrm{mPa\cdotp s}$ at $313.2\;\mathrm{K}$ vs $7.1\;\mathrm{mPa\cdotp s}$ at $413.2\;\mathrm{K}$, both at $1\;\mathrm{atm}$)\cite{Qiao-2011}. Indeed, it has been previously shown that ion packing and hence, the capacitance, shows little changes with increasing the temperature from $298\;\mathrm{K}$ to $400\;\mathrm{K}$ \cite{Feng-2013, Vatamanu-2014}. Thus, the qualitative conclusions drawn in this paper can be extrapolated to systems at lower temperatures. The IL, composed of $4194$ ions pairs, is represented by means of a coarse-grained model developed by Roy and Maroncelli \cite{Roy-2010}, in which the cations and anions are respectively characterized by $3$ and $1$ interaction sites. The potential parameters used for the carbon atoms of the electrodes were $\sigma_{C} = 3.37\;\mathrm{\AA{}}$ and $\varepsilon_{C} = 0.23\;\mathrm{kJ/mol}$ following Cole and Klein \cite{Cole-1983}. The LJ parameters for dissimilar atoms were computed using the Lorentz-Berthelot mixing rules. 

\begin{figure*}[ht!]
\includegraphics[width=0.7\textwidth]{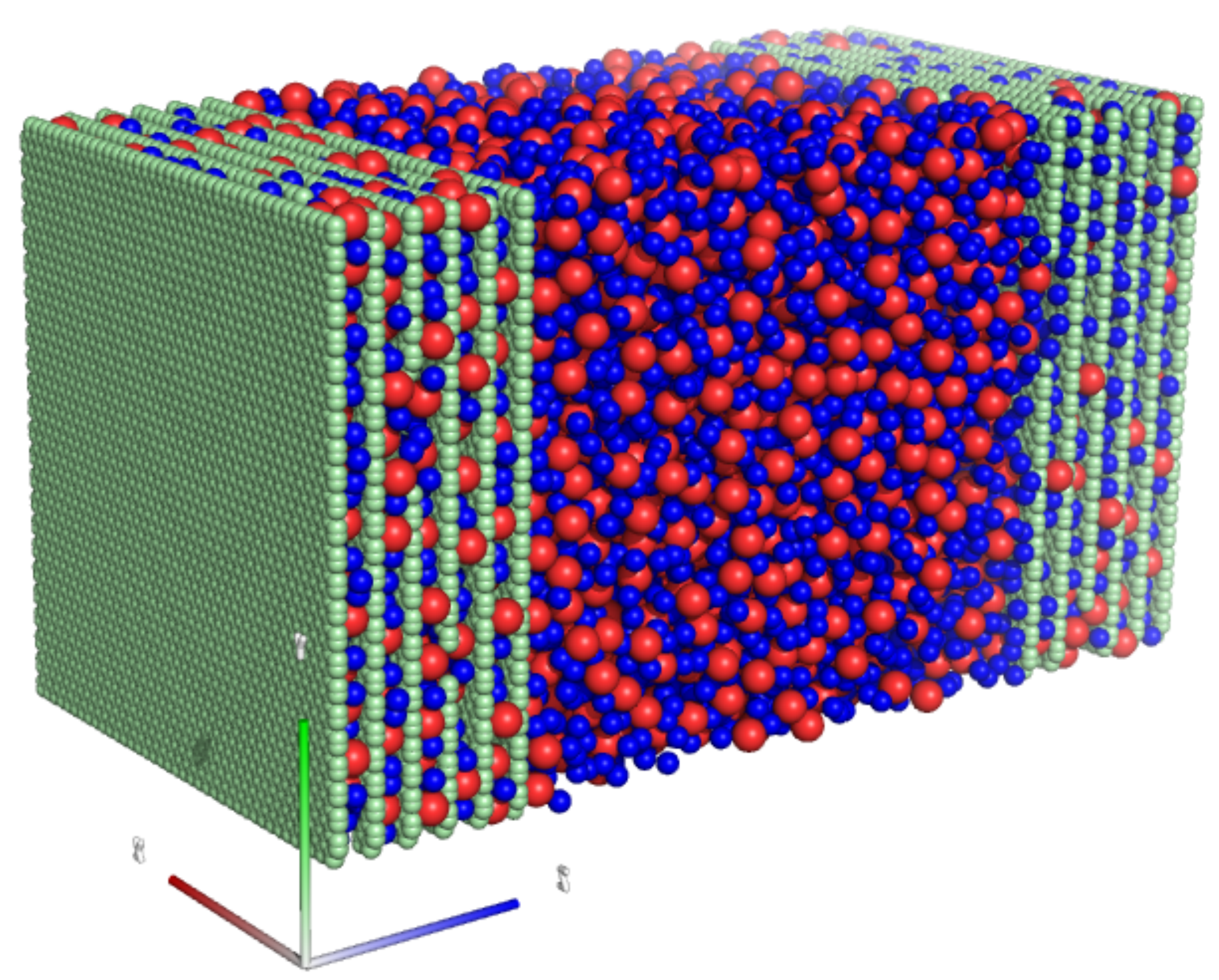}
\caption{Snapshot of the simulation box: BMIM$^+$ cations (blue) and PF$_6^-$ anions (red) confined between two graphene-based electrodes (green) with a pore size of $7\;\mathrm{\AA}$. A coarse-grained model is used for the electrolyte. \revision{\copyright Trinidad Mendez-Morales, available at \url{http://dx.doi.org/10.6084/m9.figshare.5746737.v1} under a CC-BY4.0 license.}}
\label{fig:snapshot}
\end{figure*}

Each of the electrodes is composed of 21245 carbon atoms distributed into \revision{six} fixed graphene layers. The five inner \revision{layers} are randomly perforated with round $10\;\mathrm{\AA{}}$-radius holes (2 per \revision{layer}) in order to allow the diffusion of the ions in the z direction (perpendicular to the surface) and the entrance of the liquid into the porous electrodes with fast charging and discharging rates \cite{Hu-2014}. With the aim of analyzing the effect of the \revision{slit-}pore \revision{width} on the capacitive properties of the system, two distances between consecutive graphene \revision{layers} ($7\;\mathrm{\AA{}}$ and $10\;\mathrm{\AA{}}$) were considered. The electrolyte is confined between the two electrodes as shown in Fig. \ref{fig:snapshot}. Periodic boundary conditions are applied along two directions, $x$ and $y$, and electrostatic interactions are calculated employing a two-dimensional Ewald summation \cite{Reed-2007}.

In order to obtain a realistic description of the properties of these devices, the simulations were carried out by holding the electrodes at constant potential \cite{Merlet-2012}. The model, developed by Reed \textit{et al.}\cite{Reed-2007} based on the previous work of Siepmann and Sprik \cite{Siepmann-1995}, uses fluctuating charges on each carbon atom that are calculated at each time step by requiring that the potential on every atom is constant and equal to the specified electrode potential. To achieve this polarization of the electrodes, the charges on the electrode atoms are represented by Gaussian distributions centered on the carbons, and they are calculated by using a conjugate gradient method to minimize the expression

\begin{equation}
U=\sum_{i}{q_{i}(t)}\left[ \frac{V_{i}(\{q_{j}(t)\})}{2}+\frac{q_{i}(t)}{\sqrt{2\pi}\kappa}-V^{0}\right]
\end{equation}

where $q_{i}$ is the charge on an electrode atom $i$, $V_{i}(\{q_{j}(t)\})$ is the potential felt by an electrode atom at position $i$ due to the other electrode atoms and all the ions in the electrolyte, $\kappa$ is the width of the Gaussian distributions ($\kappa = 0.5055\;\mathrm{\AA}$ in our case), and $V^{0}$ is the potential applied to the electrode.

The systems were first equilibrated at constant charge during $500\;\mathrm{ps}$, where the charge of the electrodes was fixed at $0$. Then we performed a second equilibration of $500\;\mathrm{ps}$, in which the charges of the carbon atoms were set to $+0.01e$ and $-0.01e$ in the left and the right electrodes, respectively. We then carried out production runs (with a time step $dt=2\;\mathrm{fs}$) for around $1.3\;\mathrm{ns}$ using the constant applied potential method. Since this applied potential can not be chosen by calculating the Poisson potential across the cell for the constant charge runs \cite{Salanne-2015}, we performed several short simulations with different starting potentials and we chose the one for which the total charge on the electrodes showed to be nearly constant: This yielded values of $\Delta \Psi =$1.8~V and 2.7~V for the applied potential for the $7\;\mathrm{\AA{}}$ and $10\;\mathrm{\AA{}}$ pore sizes, respectively. All the comparisons are thus performed for almost similar electrode charges, which is more meaningful since the ionic \revision{charge excess is then identical}.  


\section{Results and discussion}


\subsection{Capacitance}
The gravimetric capacitance of the simulated supercapacitor is given by

\begin{equation}
C_m=\frac{1}{m}\frac{\langle Q_+\rangle}{\Delta \Psi}
\end{equation}

\noindent where $\langle Q_+\rangle$ is the average charge of the positive electrode along the full simulation and $m$ is its corresponding mass. However, in order to compare our results to experimental data, it would be necessary to determine single electrode capacitances, which is difficult because we cannot calculate the Poisson potential across the cell for such complex pore geometries~\cite{Salanne-2015}. Nevertheless, our previous estimations in the case of CDC electrodes showed that for such a relatively simple IL (in particular with a relative symmetry for the ionic sizes), the two electrodes had almost similar capacitances. We will therefore assume that $C_m^+\approx C_m^-\approx 2C_m$ for comparison purposes. We then obtain individual capacitances of 96 and 64~F/g for the pore sizes of 7 and 10~\AA, respectively. This falls well within the range of measured values in non-activated graphene-based supercapacitors (i.e. between 45 and 130 F/g~\cite{han2014b,tsai2013a} as discussed above). 

\begin{figure*}[ht!]
\includegraphics[width=0.7\textwidth]{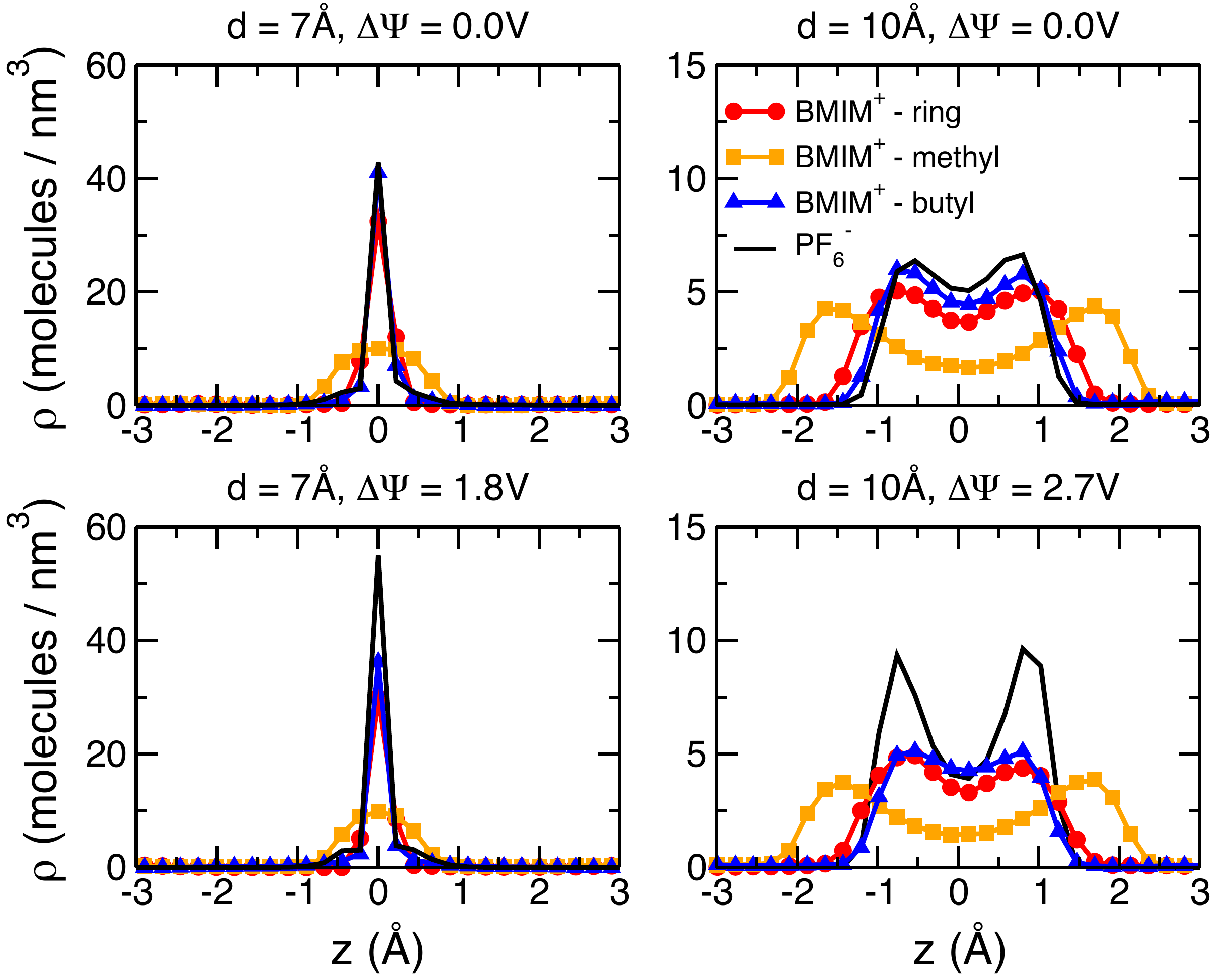}
\caption{Number density profiles inside the central pore of a neutral (top) and positive (bottom) electrodes, in the direction perpendicular to the electrodes (the $z$ coordinate is centered in the middle of the pore). $d$ is the pore size (7~\AA~in the left column and 10~\AA~in the right one) and $\Delta \Psi$ the applied potential. \revision{\copyright  Trinidad Mendez-Morales and Mathieu Salanne, available at \url{http://dx.doi.org/10.6084/m9.figshare.5746737.v1} under a CC-BY4.0 license.}}
\label{fig:dens_prof}
\end{figure*}

\subsection{Layering inside the pores}
In previous studies performed on single slit-pore electrodes, an increase of the capacitance was observed when decreasing the pore size down to the ionic dimensions~\cite{Jiang-2011, Feng-2011, Vatamanu-2015, Bedrov-2013}. This result was attributed to the transition from a monolayer to a bilayer structure for the adsorbed IL. To check whether the differences in the capacitance we observe between the two pore sizes arise from similar effects, we analyzed the structure of the IL inside each of the pores. Number densities ($\mathrm{\AA^{-3}}$) of the ions along the z-direction normal to the electrodes were calculated in both cases and the profiles obtained in the neutral and positive electrodes are shown in Fig. \ref{fig:dens_prof}. When the width of the pore is $7\;\mathrm{\AA}$, we observe that the strong confinement of the walls forces the cations and the anions to occupy a single narrow layer in the center of the pores of both electrodes (Fig. \ref{fig:dens_prof}a), with only the smaller methyl groups of cations approaching closer to the surface. Since the neutral electrodes (i.e. for an applied potential of 0~V) are already wet by the IL, applying a potential difference leads to a greater intercalation of counter-ions inside the electrodes and the expulsion of some co-ions from the pores, whereas the total number of ions per unit volume of the nanopores remains nearly the same. In this case, the IL is organized in a tight configuration that impedes an increase in ion density upon increasing the applied potential, due to which the surface charge accumulation takes place by means of a swapping mechanism. In this case, even though the co-ions are in direct contact with the charged surface due to the monolayer confinement, we did not observe a regime of complete intra-pore co-ion depletion.  

On the contrary, in the widest pore (10~\AA), the distribution of the ions is markedly different (Fig. \ref{fig:dens_prof}b). In this case, when the system is not charged the ions accumulate in two separate layers near the pore walls, since the greater available space in the pores allows larger fluctuations in the structural distribution of the ions. It must be noted that the total number of ions adsorbed into the electrodes (given by the integral of the density) increases only slightly with the  increasing pore size; which indicates that the smaller the pores, the more tightly packed the ions \cite{Kondrat-2011A}. The bilayer structure therefore does not consist of two monolayers. Once we apply a potential difference the ion population inside the positive electrode changes in the same way as noted above (ion exchange) but now, there is also a rearrangement of anions that tend to approach closer to the surface. The minimum between the two anionic density peaks is thus more pronounced than in the neutral electrode.

\begin{figure}[ht!]
\includegraphics[width=1.0\columnwidth]{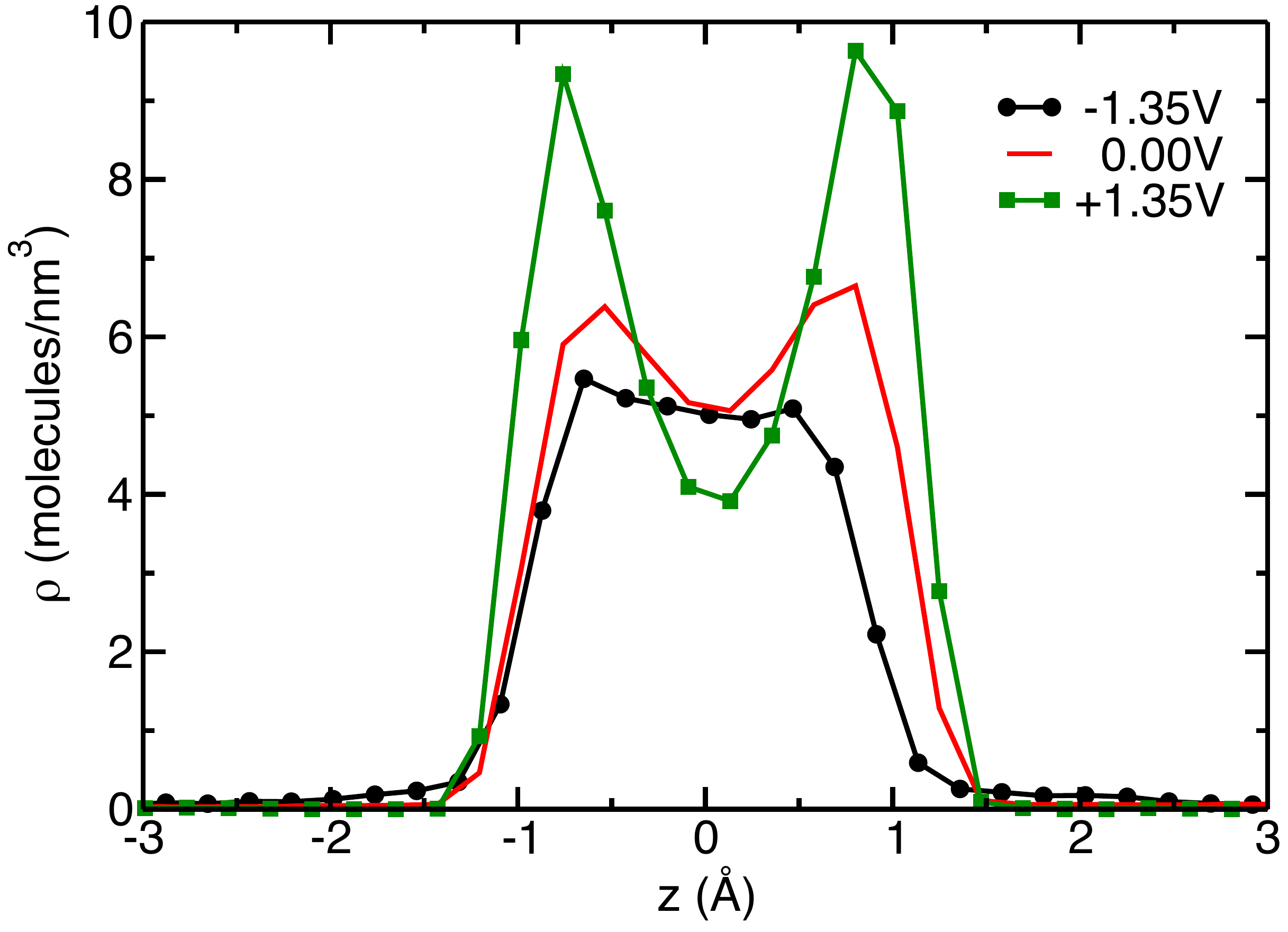}
\caption{Number density profiles for the anions adsorbed in the central pore of the 10~\AA\ pore size electrode, for applied potentials of 0~V (red) and 2.7~V (black: negative electrode, green: positive electrode). \revision{\copyright  Trinidad Mendez-Morales and Mathieu Salanne, available at \url{http://dx.doi.org/10.6084/m9.figshare.5746737.v1} under a CC-BY4.0 license.}}
\label{fig:dens_negative}
\end{figure}

In the negative electrode, the charging mechanism for the 7~\AA\ pores is almost symmetrical with respect to the positive one. This is not the case for the 10~\AA\ pores where we observe that almost all the anions concentrate in the middle of the pores, thus losing their bilayer structure, as shown on Fig. \ref{fig:dens_negative}. This is reminiscent of the structure which was reported for the same IL adsorbed inside CDC electrodes~\cite{Merlet-2012}, in which the co-ions tend to occupy the center of the pores, but also to the case of an IL adsorbed in the interlayer of negatively charged clays~\cite{duarte2014a}. Again, we observe that the structure of this monolayer of anions differs markedly from the one observed in the 7~\AA\ pores since the corresponding peak in the density is much wider and less high.  


\subsection{Decoordination inside the pores}
\begin{table}
\begin{center}
\begin{tabular}{lcc}
 \hline
{} & \textbf{Anion-Cation} & \textbf{Cation-Anion}  \\
\hline
{} & {posit. elec.} & {negat. elec.}\\
\hline
CDC (1.0 V) & 2.6 & 2.5 \\
$d=7 $\AA\ (0.0 V) & 2.9 & 2.7 \\
$d=7 $\AA\  (1.8 V) & 2.3 & 2.2 \\
$d=10 $\AA\  (2.7 V) & 2.3 & 2.3 \\
\hline
\end{tabular}
\end{center}
\caption{Coordination numbers of cations surrounding anions in the positive electrode (left), and anions surrounding cations in the negative electrode (right). The results in CDCs were extracted from reference \citenum{Merlet-2013}. \revision{For comparison, the bulk values for A-C and C-A are equal to 4.8.}}
\label{table:coord_numbers}
\end{table}

As it was previously reported \cite{Simon-2008}, a partial desolvation of the ions occurs under confinement (or rather a decoordination in solvent-free ILs). The associated loss of stability is compensated by the charge of the electrodes and it has been shown to be directly related with the maximum capacitance for pores smaller than the solvated ion size \cite{Simon-2006, Simon-2013}. In Table \ref{table:coord_numbers} we can observe that the coordination numbers of cations around anions (left) and {\it vice versa} (right) decreased from $4.8$ in the bulk to less than $3$ inside the nanopores. The coordination numbers in each electrode were calculated by averaging the number of co-ions that were closer to a central counter-ion than $7.5\;\mathrm{\AA}$, which is the first minimum of the anion-cation radial distribution function (RDF) in the bulk. When compared with CDCs\cite{Merlet-2013}, the ions confined in perforated graphene-based electrodes are, on average, less coordinated. It must be noted that although the distributions of the coordination numbers are centered around $2$ in both cases, in the CDC case occurrences of up to $7$ are observed, which is due to the presence of different adsorption sites\cite{Merlet-2013}. As expected, a greater desolvation of the counter-ions is obtained inside charged electrodes. However, we did not register an influence of the pore size on the coordination numbers, because even though the ion density is greater in the $7\;\mathrm{\AA}$-system, the 2D-solvation of the ions is characterized by larger lateral distances between the co-ions remaining inside the pores and the counter-ions introduced into them \cite{Qiao-2011}.   

\subsection{Partial breaking of the Coulombic ordering}
\begin{figure}
\includegraphics[width=\columnwidth]{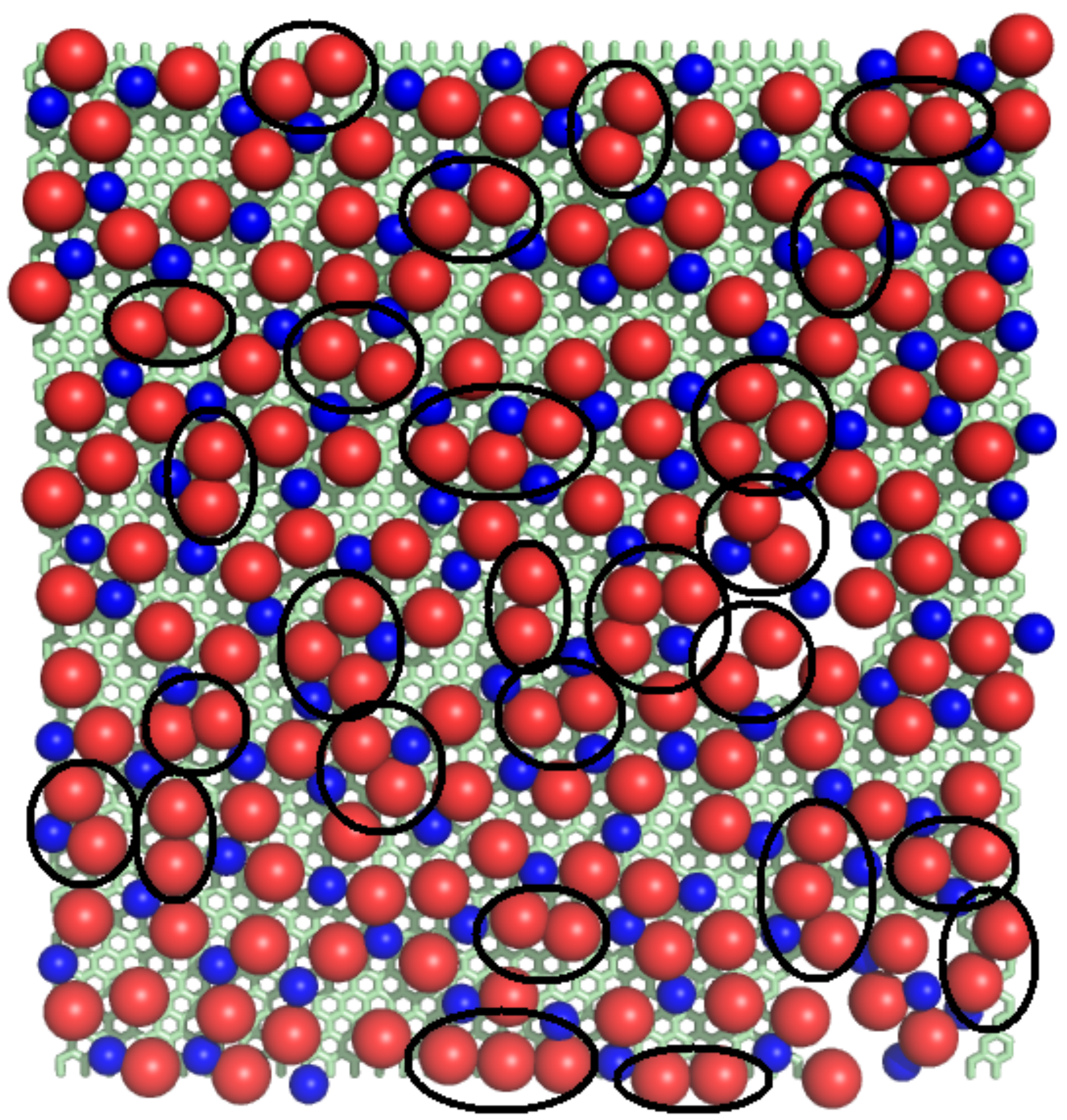}
\caption{Illustration of the breaking of the Coulombic ordering. Snapshot of an adsorbed layer in the positively charged 7~\AA\ pore (blue: imidazolium ring of the cations, red: anions). Pairs (and triplets) of co-ions are surrounded by black ovals. \revision{\copyright Trinidad Mendez-Morales, available at \url{http://dx.doi.org/10.6084/m9.figshare.5746737.v1} under a CC-BY4.0 license.}}
\label{fig:sup_ionic}
\end{figure}

Our results suggest the stabilization of the partially desolvated counter-ions inside the nanopores by means of a charge compensation from the electrodes, which could be explained by the picture reported by Kondrat and Kornyshev \cite{Kondrat-2011, Kondrat-2011A} of an exponential screening of the electrostatic interactions in a slit-like metallic pore due to the image charges. This superionic state breaks the Coulombic ordering characteristic of ILs and allows more ions of the same sign to occupy the pore, which leads to a high concentration (mainly limited by steric repulsions) of partially desolvated counter-ions that can be in contact with the charged walls.  The screening of the electrostatic interactions is expected to sharply increase in narrower pores, thus yielding a significant improvement of the capacitive properties. This theoretical finding has also recently been confirmed experimentally by performing X-ray scattering experiments to resolve the structure of the EMI-TFSI ionic liquid confined inside electrified nanoporous carbons~\cite{futamura2017a}. To analyze the presence of such a dense ionic state in our simulation, we computed the ratio of the number of anion-anion (cation-cation) pairs in the positive (negative) electrode to the number of anions (cations) in that electrode. We considered the formation of a pair of ions with similar charges if they are located closer than $5.8\;\mathrm{\AA}$, which is the distance at which the anion-anion (or cation-cation) and anion-cation RDFs (both in the bulk) intersect each other (for cations the interaction center corresponding to the imidazolium ring was used to perform this analysis). A set of such pairs is highlighted on a snapshot of the central adsorbed layer in the positively charged 7~\AA\ pore on Figure \ref{fig:sup_ionic}. We can observe that the fractions increase with the confinement, switching from 0.049 to 0.109 for anions in the positive electrode and from 0.099 to 0.141 for cations in the negative one. Albeit on a different system, these numbers are in good agreement with the ones reported in the experimental study of Futamura {\it et al.}~\cite{futamura2017a}, thus providing further evidence for the existence of the superionic state for ionic liquids adsorbed inside ultranarrow electrified nanopores.  

\begin{figure*}
\includegraphics[width=\textwidth]{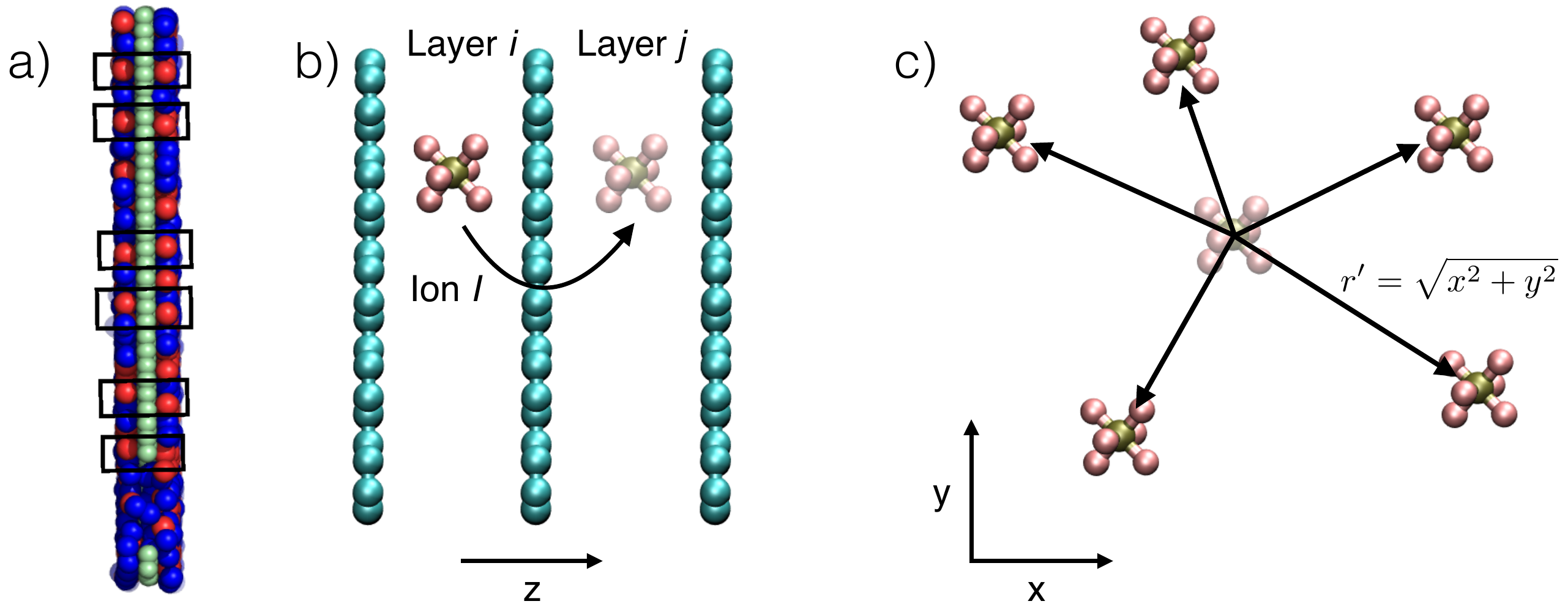}
\caption{a) Snapshot of a a portion of the system representing two adjacent pores separated by a graphene sheet for the simulation performed with $d=7$\AA\ and $\Delta \Psi=$1.8~V (red: anions, blue: imidazolium rings of the cations, green: carbon atoms). b) Projection of the ion $I$ from layer $i$ to the layer $j$. c) The in-plane distances $r'$ between the projected ion and all the ions in layer $j$ are then computed. \revision{Note that panels b) and c) show all the atoms in the PF$_6^-$ anion, but a coarse-grained model was used in the simulation.} \revision{\copyright Trinidad Mendez-Morales and Mathieu Salanne, available at \url{http://dx.doi.org/10.6084/m9.figshare.5746737.v1} under a CC-BY4.0 license.} }
\label{fig:correlations1}
\end{figure*}
  
\subsection{Ion-ion correlations from pore to pore}

The results described above concerning the formation of mono/bilayer(s), decoordination and the formation of pairs of anions or cations are in agreement with the previous literature on single slit pore simulations. However, our original simulation setup allows the study ionic interactions from one pore to another. In a recent work, a 
templating effect of graphene was recently reported by Chae {\it et al.}~\cite{chae2017a}. They showed that the atomic arrangements of ZnO nanocrystals nucleated on a graphene plane  
exhibit a close match with those of a substrates bound to the other side of the
graphene layer. In our case a simple visualization of the atomic positions in two adjacent pores (Figure \ref{fig:correlations1}a) suggests that similar mechanisms may be at play.
In order to investigate the existence of such correlations between ions adsorbed in adjacent pores, we
compute the two-dimensional radial distribution function (RDF) between an ion 
within a given interlayer and the ions in a neighbouring one. As illustrated on Figure \ref{fig:correlations1}b and c, we project an anion $I$ with coordinates $(x_I,y_I,z_I)$ from the layer $i$ to the layer $j$. The coordinates of this projection are then $(x_I,y_I,z_I+d)$ where $d$ is the pore size. We then define an in-plane distance between this projected anions and any anion $J$ present inside layer $j$ as

\begin{equation}
r'_{IJ}=\sqrt{(x_I-x_J)^2+(y_I-y_J)^2}
\end{equation}

\noindent We then perform an ensemble average to determine the in-plane anion-anion RDF between layers $i$ and $j$ ($g_{ij}$):

\begin{equation}
g_{ij}(r')=\frac{S}{N_I N_J 2 \pi r' dr'}\langle \sum_{I \in  i} \sum_{J \in  j} \delta(r'-r'_{IJ})\rangle
\end{equation}  
\noindent where $N_I$ and $N_J$ are the number of anions inside layers $i$ and $j$, respectively, $S$ is the lateral surface of the simulation cell (along $x$ and $y$ coordinates) and $dr'$ is the precision chosen for binning the Kronecker function $\delta$. Similar RDFs can be defined for cation-cation and anion-cation pairs.

\begin{figure}
\includegraphics[width=\columnwidth]{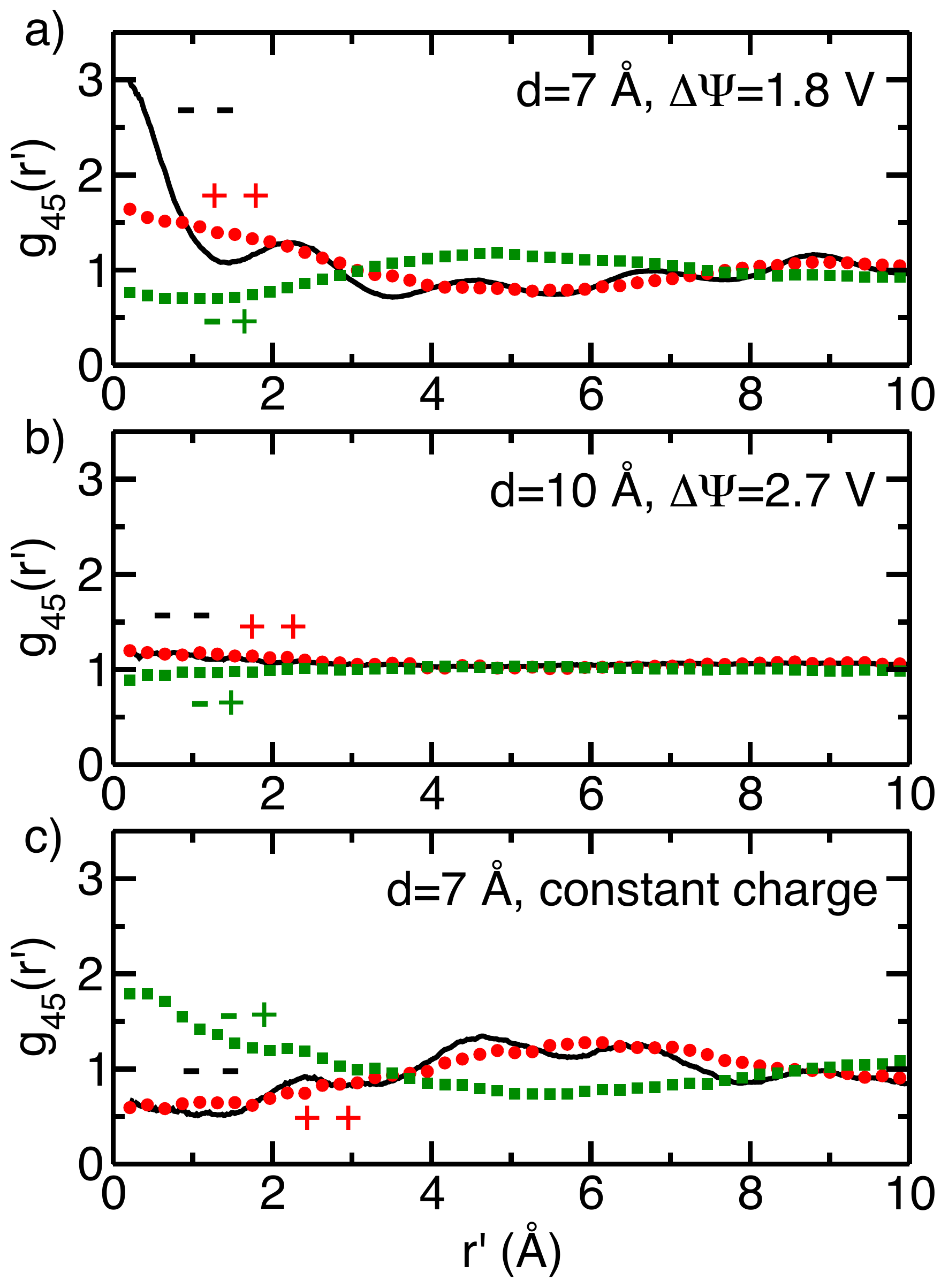}
\caption{In-plane radial distribution functions between an ion 
within the fourth layer of adsorbed liquid and the ions in the fifth one (black: anion-anion, red: cation-cation and green: anion-cation). Panels \revision{a) and b)}: A constant potential is applied to the electrode, allowing the surface charge to fluctuate in response to the electrolyte (\revision{a}: $d=7$\AA\, $\Delta\Psi$=1.8~V, \revision{b}: $d=10$\AA\, $\Delta\Psi$=2.7~V). \revision{Panel c)}: The partial charges on the carbon atoms are held constant and equal to the average charge for the 1.8~V applied potential simulation ($d=7$\AA). \revision{\copyright Trinidad Mendez-Morales and Mathieu Salanne, available at \url{http://dx.doi.org/10.6084/m9.figshare.5746737.v1} under a CC-BY4.0 license.}}
\label{fig:correlations2}
\end{figure}

Examples of such functions are provided for various simulation setups in Figure  
\ref{fig:correlations2} for the correlations between the fourth and fifth layers of liquid (similar results were obtained for all the adjacent layers). Firstly, very large differences are observed between the electrified pores with a size of 7~\AA\ (top) and 10~\AA\ (middle). Indeed, for the former we observe a strong peak at $r$ = 0~\AA\  for the anion-anion, which shows that anions
have a preference for adsorbing in front of other anions. On the contrary, the cation population is depleted
in such sites as can be seen from the low value taken by the anion-cation RDF. The cation-cation RDF is qualitatively similar, albeit its shape is smoother due to the greater flexibility that is allowed to the BMIM$^+$ in the coarse-grained model.  The structure of each fluid layer is therefore clearly
correlated with that in the neighbouring ones across graphene sheets, 
with fluctuations due to the temperature and the presence
of holes inside the graphene plane.

In the 10~\AA\ pores, these correlations are almost completely lost. A small preference for like-charge ions is still observed since the anion-anion and cation-cation RDFs slightly rise above 1 for short distances, but this effect is almost not noticeable. It is therefore necessary to have a well-defined single layer of ions to observe the effect.

In order to check whether these correlations arise from the
image charges induced by the ions themselves on the surface of the carbon, we performed an additional simulation for the 7~\AA\ pores where the
average charge of the plane was equally shared between all the carbon atoms. \revision{This} 
situation \revision{is} often called ``constant charge'' simulation, although it does not
correspond to a case relevant to electrochemical simulations. As shown in the
corresponding RDFs (bottom panel of Figure \ref{fig:correlations2}),
the correlations are then lost. On the contrary, Coulombic ordering is recovered
and cations have a slight tendency to lie in front of anions from adjacent
\revision{layers}. This shows that the charge localization on the graphene surface plays a key role and that constant charge simulations should be avoided when simulating complex porous carbons in which the surface is accessible from the two sides.

\section{Conclusions}

In summary, BMIM-PF$_6$ structure in nanoporous graphene-based electrodes was analyzed by means of MD simulations using the constant applied potential method to simulate the electrode materials. Both anions and cations were significantly desolvated inside the nanopores regardless of their width, but the way they are accommodated varied from a single narrow layer to a bilayer configuration when increasing the pore size from $7\;\mathrm{\AA}$ to $10\;\mathrm{\AA}$. Additionally, the ions were found to be more densely packed in the smaller pores, forming a single monolayer while they arrange in two layers when they have access to a larger volume. The charging of the supercapacitor is then allowed by an efficient screening of the Coulombic interactions between the ions by the image charges. This superionic state previously predicted by Kondrat and Kornyshev was confirmed by a more marked formation of counter-ion/counter-ion pairs under a strong confinement.  

The main specificity of the presence of a single graphene sheet between adjacent layers of liquids is that it allows the presence of structural correlations between ions of the same sign belonging to different pores; that is, the distribution of the ions in one layer is strongly influenced by the conformation that the ions adopt inside the other layers due to the image forces. \revision{ It is worth noting that in a former DFT study, it was shown that two BF$_4^-$ ions adsorbing on the same site on both sides of a graphene layer were more stable than those on different sites~\cite{kunisada2010a}. Here} the correlation \revision{effects seem to be enhanced by a strong confinement since they are} much more remarkable in the thinner pores, and almost absent when a bilayer of liquid is allowed to form inside the electrode. We proved that here again, the  strong localization of image charges on the graphene surface plays a key role: When assigning similar constant charges to the carbon atoms (with the same average as in the constant potential simulation), the correlations are reversed and the usual Coulombic ordering is recovered. These results open further questions for the development of efficient graphene-based supercapacitors. For example, it will be of great interest to determine whether these correlations remain present in \revision{realistic models}, where restacking of the graphene layers is likely to happen. Much of the applications involve organic electrolytes~\cite{burt2016a} or solvate ionic liquids~\cite{coles2017a}, so that it will also be of importance to determine if the presence of a solvent mitigates the importance of the ion-ion correlations across the graphene layers.


\begin{acknowledgments}

This work was supported by the French National Research Agency (Labex STORE-EX,  Grant No. ANR-10-LABX-0076 and ANR SELFIE, Grant No. ANR-17-ERC2-0028) and by Defi CNRS INPHYNITI 2015-2016 (SIMELEC). We acknowledge support from EoCoE, a project funded by the European Union Contract No. H2020-EINFRA-2015-1-676629, from the DSM-�nergie programme of CEA and from the Eurotalent programme. We are grateful for the computing resources on OCCIGEN (CINES, French National HPC) and CURIE (TGCC, French National HPC) obtained through the project x2016096728.

\end{acknowledgments}



\end{document}